\documentclass[prd,nofootinbib,aps,twocolumn,floats,floatfix,
superscriptaddress,amsmath,amssymb,secnumarabic]{revtex4-1}
\usepackage{natbib}
\usepackage{moreverb}
\usepackage{graphicx}
\usepackage{amsmath}
\usepackage{amssymb}
\usepackage{float}
\usepackage{slashed}
\usepackage{color}
\usepackage[utf8]{inputenc}
\usepackage{hyperref}

\newcommand{\be}{\begin{equation}}
\newcommand{\ee}{\end{equation}}
\newcommand{\ba}{\begin{array}}
\newcommand{\ea}{\end{array}}
\newcommand{\bea}{\begin{eqnarray}}
\newcommand{\eea}{\end{eqnarray}}

\begin{document}

\title{Constraining Dwarf Spheroidal Dark Matter Halos With The Galactic Center Excess}
\author{Jeremie Choquette\footnote{jeremie.choquette@physics.mcgill.ca}}
\affiliation{Department of Physics, McGill University,
3600 Rue University, Montr\'eal, Qu\'ebec, Canada H3A 2T8}

\begin{abstract}
If the gamma-ray excess from the galactic center reported by Fermi-LAT is a signal from annihilating dark matter, one must question why a similar excess has not been observed in dwarf spheroidal galaxies. We use this observation to place constraints on the density profile of dwarf spheroidal galaxies under the assumption that the galactic center excess is in fact a signal from annihilating dark matter. We place constraints on the generalized NFW parameter $\gamma$ and the Einasto profile parameter $\alpha$ which control the logarithmic slope of the inner regions of the halo's density profile. We determine that under these assumptions the galactic center excess is inconsistent with the standard NFW profile (and other `cuspy' profiles) for dwarf spheroidal galaxies , but is consistent with observations of cored dwarf galaxy profiles. Specifically, we find that dwarf spheroidal profiles must be less cuspy than that of the Milky Way. Models of dark matter which self-interacts through a light mediator can achieve this.

\end{abstract}
\maketitle

\section{Introduction}

Observations by Fermi-LAT have indicated an excess of gamma-rays in the center of the Milky Way galaxy in the range of a few GeV~\cite{Goodenough:2009gk,Hooper:2010mq,Hooper:2011ti,
Abazajian:2012pn,Zhou:2014lva,Calore:2014xka,Daylan:2014rsa,FermiData,TheFermi-LAT:2015kwa}. Interpretations of the galactic center excess (GCE) differ, with likely candidates including dark matter annihilations and known astrophysical phenomena. On the astrophysical side, the spectrum and morphology of the signal from millisecond pulsars provides a good fit to the observed excess~\cite{Lee:2015fea,Bartels:2015aea,O'Leary:2015gfa}, but this would require a much greater number of millisecond pulsars than are observed or expected~\cite{Linden:2015qha,Cholis:2014lta}. The Fermi-LAT collaboration has more recently completed an analysis of the purported signal and has concluded that the morphology of the signal is more consistent with millisecond pulsars than with the dark matter interpretation~\cite{TheFermi-LAT:2017vmf,Fermi-LAT:2017yoi}. It is concluded that the dark matter interpretation is strongly disfavoured relative to other interpretations of the excess. In a recent paper, however, Haggard et al. argue that a sufficiently large population of millisecond pulsars would also imply a large population of observable low-mass X-ray binaries, limiting the contribution of millisecond pulsars to the galactic center excess to $\sim 4\%-23\%$\cite{Haggard:2017lyq}, leaving annihilating dark matter as a contender.

It is also well known that there is tension between dark matter explanations of the galactic center excess and observations of dwarf spheroidal galaxies. Dwarf spheroidal galaxies show no corresponding signal, with the constraints seeming to exclude dark matter annihilation as a viable explanation for the galactic center excess\cite{Ackermann:2015lka,Abazajian:2015raa}. The analysis of~\cite{Ackermann:2015lka} (upon which~\cite{Abazajian:2015raa} is based), however, assumes a Navarro-Frenk-White (NFW) profile for the dwarf spheroidals. The NFW profile has a sharp cusp at the center, leading to an enhanced signal relative to more `cored' dark matter distributions. We consider two profiles here: the generalized NFW profile and the Einasto profile, defined in equations (\ref{eq:NFW}) and (\ref{eq:Einasto}) respectively.

The exact distribution of dark matter in dwarf galaxies is not well known, but there is a large body of evidence pointing to cored profiles (see section~\ref{sec:comparison}), or profiles with inner radii with slope smaller than the $\rho\propto r^{-1}$ predicted by cold dark matter simulations and exemplified by the NFW profile.

\begin{figure}
\includegraphics[scale=0.4]{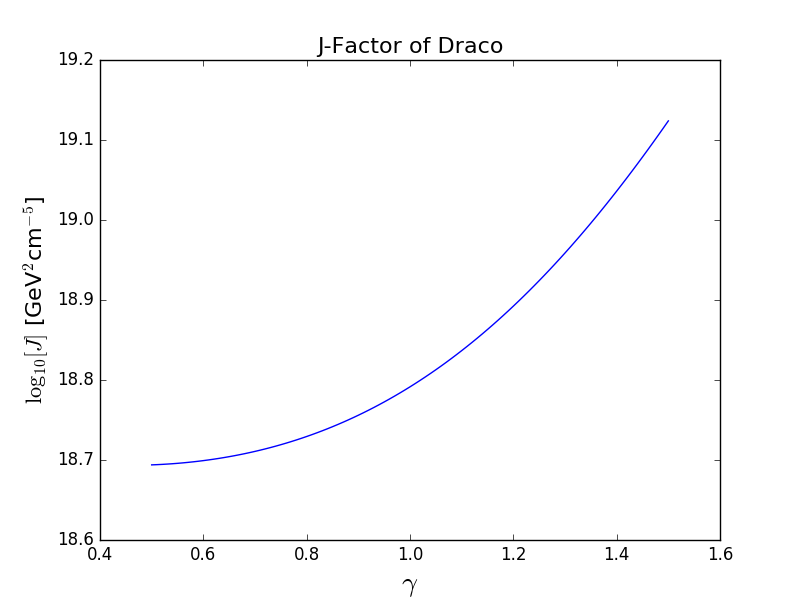}
\includegraphics[scale=0.4]{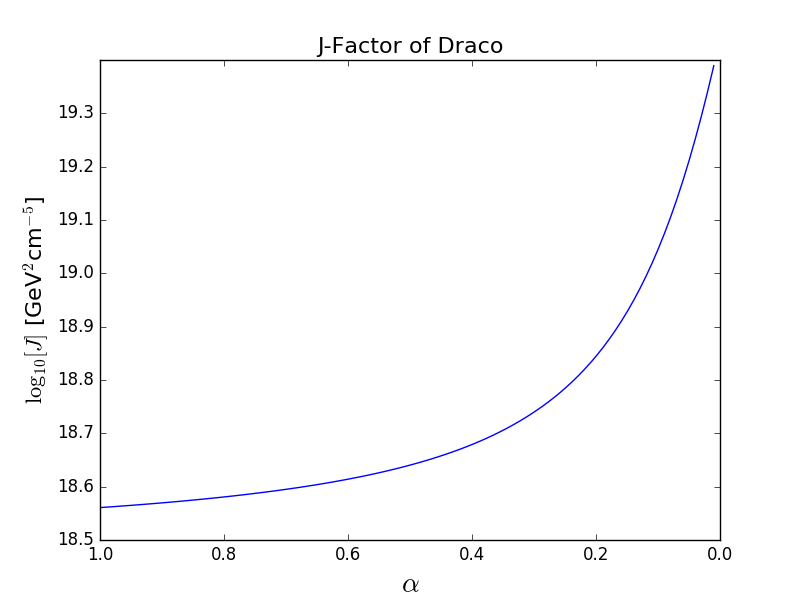}
\caption{The $J$-factor of the Draco dwarf spheroidal as a function of $\gamma$ or $\alpha$ assuming a generalized NFW (top, eq. (\ref{eq:NFW})) or Einasto profile (bottom, eq. (\ref{eq:Einasto})) defined in equations (\ref{eq:NFW}) and (\ref{eq:Einasto}) below. These parameters control the cuspiness of the NFW and Einasto profiles (respectively). The methodology is explained in section~\ref{sec:dwarfs}.}
\label{fig:DracoJ}
\end{figure}

The logarithmic slope of the inner dark matter halo can have a significant impact on its $J$-factor, a measure of the rate of dark matter annihilations within the halo. 
We can see how the parameters $\gamma$ and $\alpha$ (the parameters controlling the cuspiness for the NFW and Einasto profiles respectively, as explained below) alter the $J$-factor of Draco, for example, in figure~\ref{fig:DracoJ}. Although these differences may not seem tremendously large, the tension between the dwarf galaxy observations and the GCE is moderate, and these differences can be enough erase it entirely.

It also follows, therefore, that if the GCE signal were assumed to indeed originate from dark matter annihilations, constraints could be placed on the central slope of the dark matter profiles of the dwarf spheroidals. In section~\ref{sec:sim} we simulate the GCE signal from dark matter to find best fit values for the dark matter mass and annihilation cross section. In section~\ref{sec:dwarfs} we use these adopted values to place limits on the parameters $\gamma$ and $\alpha$ which control how cuspy the dwarf spheroidals are. In section~\ref{sec:comparison} we compare these values to those found through observation of dwarf spheroidals and simulations of cold dark matter (CDM) halos. In section~\ref{sec:beyondcdm} we discuss the implications for the CDM paradigm, should the GCE prove to indeed be a signal from annihilating dark matter.

\section{Simulation of Signal}
\label{sec:sim}

It has been shown that the observed gamma ray excess is well fit by models of annihilating dark matter in which the dark matter predominantly annihilates to $b\bar{b}$. The signal, however, consists of multiple components: the prompt gamma rays (from the $b$ decay products), inverse Compton scattering (ICS, caused by the upscattering of starlight and CMB photons by the $e^+/e^-$ produced as $b$ decay products) and a small amount of bremsstrahlung radiation (also from the decay products).  These three sources combine to produce the total signal.

The prompt signal is easiest to compute numerically, as it depends only on the $J$-factor and average spectrum from a single annihilation, taken from PPPC 4~\cite{Cirelli:2010xx,Ciafaloni:2010ti}:
\begin{align}
\frac{d\Phi_{\rm prompt}}{dE}&=\frac{\langle\sigma |v|\rangle}{8\pi m_\chi^2}\frac{dN_\gamma}{dE}\times J,\\
\label{eq:prompt}
J&=\int_{\Delta \Omega}\int_{\rm l.o.s.} \rho^2 dld\Omega,
\end{align}
with the integral along the line of sight and angular extent of the observed system.
The $J$-factor can then be computed numerically by assuming a density profile for the dark matter halo.

One way to parametrize the cuspiness of a galaxy is through the inner slope of the profile. If we assume a generalized NFW profile:
\begin{align}
\rho(r)=\frac{\rho_s}{\left(\frac{r}{R_s}\right)^\gamma\left(1+\frac{r}{R_s}\right)^{3-\gamma}},
\label{eq:NFW}
\end{align} 
then the parameter $\gamma$ corresponds to the negative slope at $r=0$. Larger values of $\gamma$ correspond to a more cuspy profile, whereas smaller values correspond to a more cored profile. Following~\cite{Calore:2014xka} we choose a generalized NFW profile with $R_s=20$ kpc and $\rho_\odot=0.40\,\rm{GeV}\rm{cm}^{-3}$ (the local dark matter density, which for $\gamma=1$ corresponds to a scale density of $\rho_s=0.26\,\rm{GeV}\rm{cm}^{-3}$). $\gamma$ is typically taken to be somewhere on the order of $1.0-1.5$, with $\gamma=1.0$ corresponding to the classic NFW profile, but in our analysis we allow it to vary from $0.1-1.4$.

Another popular profile that is easily parametrized in terms of the inner slope is the Einasto profile:
\begin{align}
\rho(r)&\propto e^{-A\,r^{\alpha}},\\
\rho(r)&=\rho_se^{-\frac{2}{\alpha}\left(\left(\frac{r}{R_s}\right)^\alpha-1\right)}.
\label{eq:Einasto}
\end{align}
The parameter $A$ and the proportionality constant are chosen maintain the same slope and density at $R_s$ as the NFW profile.
Although the parameter $\alpha$ does not exactly correspond to the inner log slope, it does control the extent to which the profile is concentrated toward the center, with greater concentrations at smaller $\alpha$.
We therefore consider both Einasto and NFW profiles in our analysis, using $\gamma$ and $\alpha$ to control how cuspy the profile is.

For the ICS and bremsstrahlung predictions, we use simulations to account for the propagation of decay products through the Milky Way and the distribution of gas and photons.  We use the DRAGON code~\cite{Evoli:2008dv} to simulate the injection and propagation of high energy electrons from DM annihilation, and the GammaSky program to compute the ICS and bremsstrahlung contributions resulting from these cosmic rays. GammaSky is as yet unreleased, though some results have been given~\cite{DiBernardo:2012zu}. GammaSky implements GALPROP in the calculation of photon production and upscattering along the line of sight.

\begin{figure}
\includegraphics[scale=0.4]{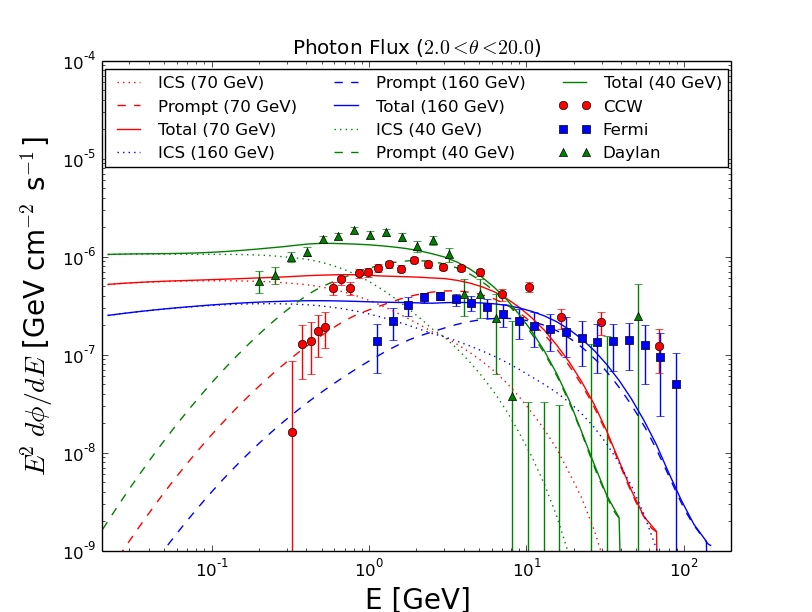}
\caption{Example of simulated GCE signal (NFW profile, $\gamma_{\rm MW}=1$) compared to that observed by~\cite{Calore:2014xka} (red),~\cite{Daylan:2014rsa} (green), and~\cite{FermiData} (blue). The simulated signal is shown for the individual best fit values in table~\ref{eq:adopted}.}
\label{Fig:spectrum}
\end{figure}

We use the model parameters --- describing the galactic magnetic field strength and shape and the galactic diffusion model used to compute the resulting inverse Compton scattering rates --- adopted in~\cite{Calore:2014xka}, labelled Model F, which is found to perform particularly well in explaining the GCE signal. We compare the results for a range of dark matter masses (20 GeV $\leq m_\chi \leq$ 200 GeV) to the GCE signals estimated in~\cite{Calore:2014xka,Daylan:2014rsa,FermiData}, as shown in~\ref{Fig:spectrum}. 

\begin{figure}
\includegraphics[scale=0.4]{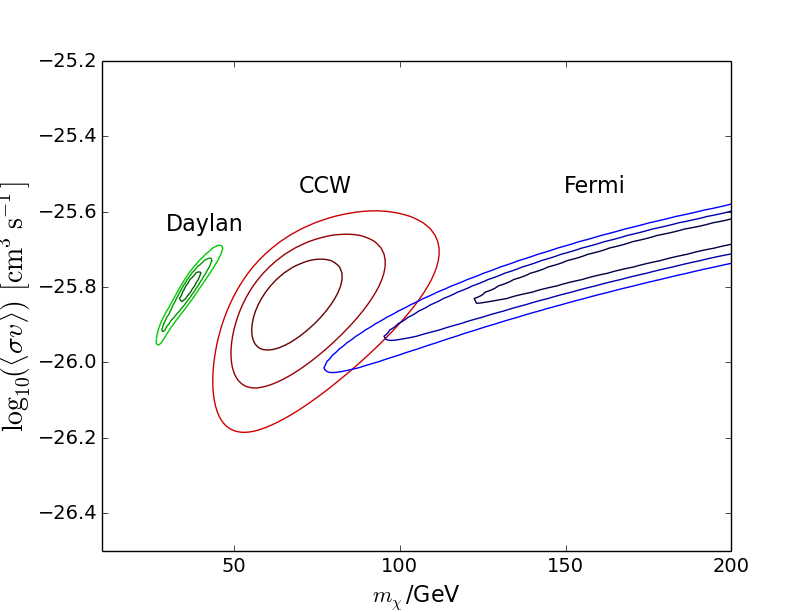}
\caption{Example of best fit $\chi^2$ $1\sigma$, $2\sigma$ and $3\sigma$ contours for~\cite{Calore:2014xka} (red),~\cite{Daylan:2014rsa} (green), and~\cite{FermiData} (blue). This example is for an NFW profile, $\gamma_{\rm MW}=1$.}
\label{Fig:contours}
\end{figure}

Figure~\ref{Fig:contours} shows the best fit regions for $\gamma_{\rm MW}=1$, showing the $1-\sigma$, $2-\sigma$, and $3-\sigma$ confidence intervals generated by minimizing the $\chi^2$ and creating contours at $\chi^2_{\rm min}+2.30,~+6.18,$ and $+11.93$. This gives us our best fit values which we will adopt when placing limits on the dwarf galaxy profiles. An example, for $\gamma_{\rm MW}=1$, is shown in table~\ref{eq:adopted}.
\begin{table}[H]
\begin{center}
\begin{tabular}{|l|cc|c}
\hline
Dataset&$\langle\sigma|v|\rangle\,[\rm{cm}^3\rm{s}^{-1}]$&$m_\chi\,[\rm{GeV}]$\\
\hline
CCW&$1.5\times10^{-26}$&$70$\\
Fermi&$1.9\times10^{-26}$&$160$\\
Daylan&$1.7\times10^{-26}$&$40$\\
\hline
\end{tabular}
\caption{Best fit values found for $\gamma_{\rm MW}=1$.}
\label{eq:adopted}
\end{center}
\end{table}

\section{The Dwarf Spheroidal J Factors}
\label{sec:dwarfs}

Given the assumption that the GCE signal is indeed the result of annihilating dark matter, our adopted values can be used to place constraints on the density profiles of dwarf spheroidal galaxies. We once again assume an NFW or Einasto profile, allowing the parameters $\gamma_{\rm dpsh}$ and $\alpha_{\rm dsph}$ to range from $0.1-1.2$ and $0.01-1.0$ respectively.

The exact halo parameters $R_s$ and $\rho_s$ of the dwarf spheroidals are not well known for either profile. Given the difficulty of measuring a large enough population of stars in the galaxies combined with the fact that they are very dark-matter dominated, stellar kinematic surveys tend to give us a view of the profiles of only the innermost regions of many dwarf spheroidals. Furthermore, these parameters themselves depend on the shape of the profile assumed; a given dwarf spheroidal will have different values for its characteristic radius and density depending on what value of $\gamma_{\rm dsph}$ or $\alpha_{\rm dsph}$ is chosen.  We therefore derive best fit parameters for individual values of $\gamma_{\rm dsph}$ and $\alpha_{\rm dsph}$ using the maximum likelihood method described in appendix~\ref{app:likelihood}, using stellar kinematic data.

With our adopted value for the annihilation cross section from the fit to the GCE data, we can find the expected signal from any individual dwarf galaxy as a function of the dark matter mass $m_\chi$ using equation (\ref{eq:prompt}). Note that we only consider the prompt signal for dwarf spheroidal galaxies as they are much cleaner environments and therefore have negligible contributions from inverse Compton scattering or bremsstrahlung radiation.

\begin{figure}
\includegraphics[scale=0.4]{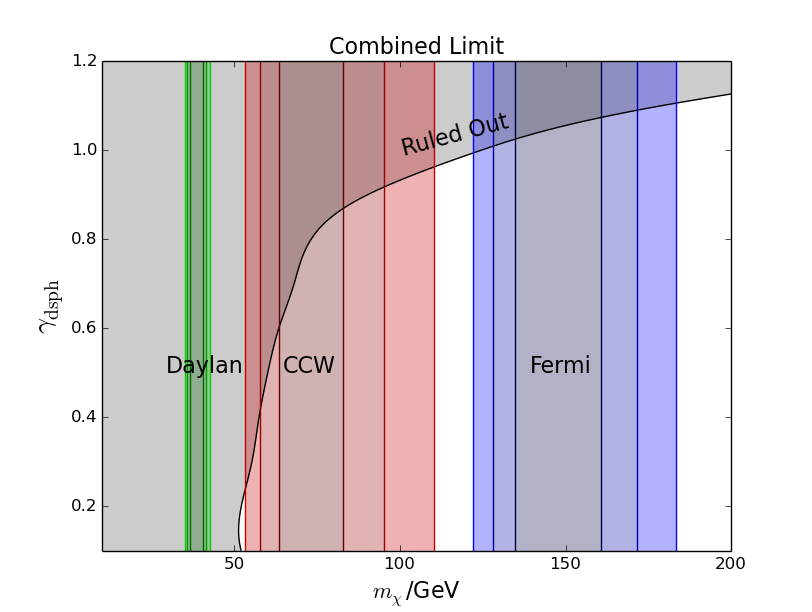}
\includegraphics[scale=0.4]{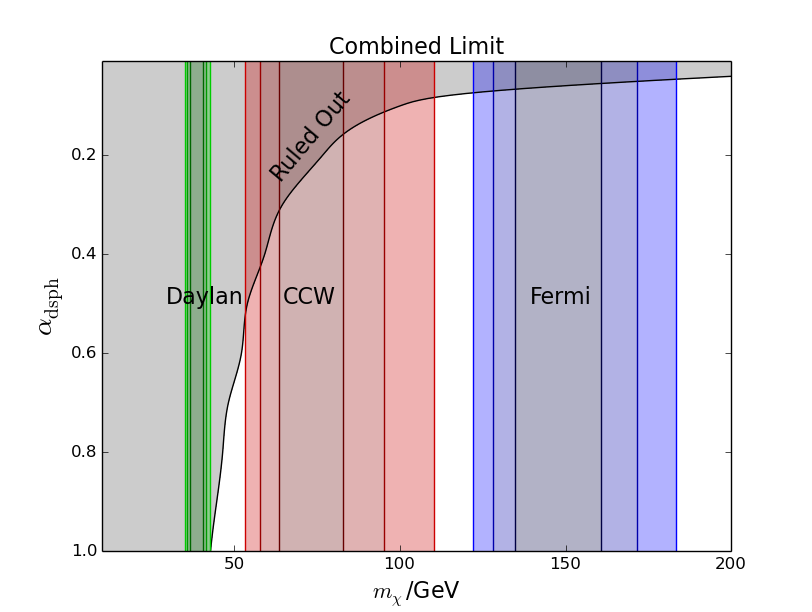}
\caption{$95\%$ C.L. constraints on $\gamma_{\rm dsph}$ and $\alpha_{\rm dsph}$ for the generalized NFW (top) and Einasto (bottom) profiles respectively. The best-fit contours for the fit to the GCE are shown in red~\cite{Calore:2014xka}, green~\cite{Daylan:2014rsa}, and blue~\cite{FermiData}. We assume $\gamma_{\rm MW}=1.0$ and $\langle\sigma|v|\rangle=1.7\times10^{-26}\,\rm{cm}^3\rm{s}^{-1}$.}
\label{fig:dwarfconstraints}
\end{figure}

The Fermi-LAT collaboration has released the upper limits on the observed flux from a large number Milky Way dwarf spheroidal galaxies based on 6 years of observation~\cite{Ackermann:2015zua}.
We compare our simulated observed flux to these reported limits, assuming an observed flux of 0 and taking their $95\%$ C.L. limit as twice the $1-\sigma$ deviation. Computing the $\chi^2$ of our simulations versus their observations, we obtain a $95\%$ C.L. constraint on the halo parameters as a function of mass by finding the contour along which $\chi^2=\chi^2_{\rm min}+6.18$. The resulting constraints are shown in figure~\ref{fig:dwarfconstraints}.

\begin{figure}
\includegraphics[scale=0.4]{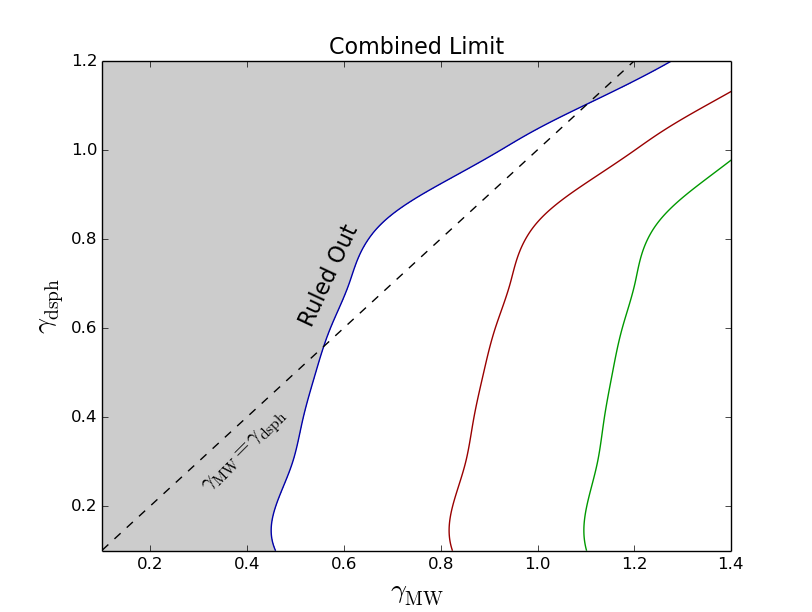}
\includegraphics[scale=0.4]{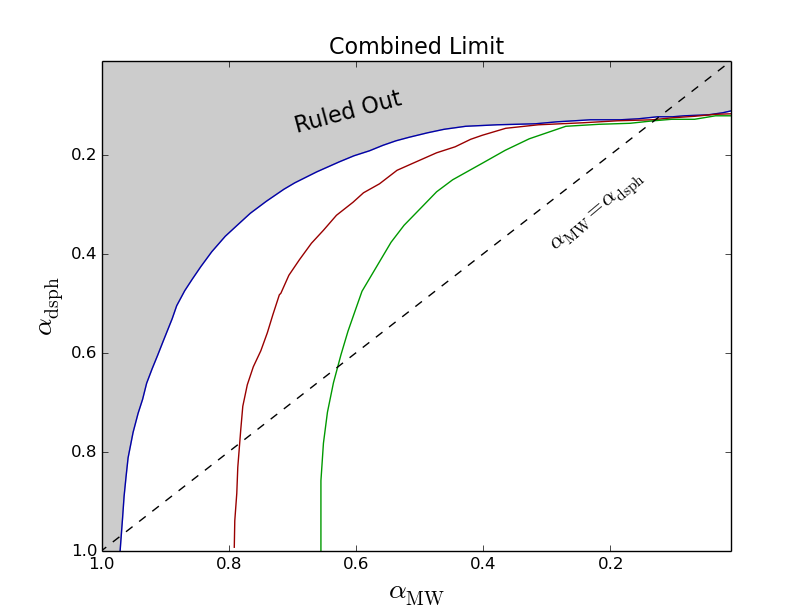}
\caption{$95\%$ C.L. constraints on $\gamma$ for both the Milky way and the Dwarf Spheroidals. The signals are calculated for the individual best-fit masses and annihilation cross sections for each of the three datasets, as shown in Figure~\ref{Fig:contours}.}
\label{fig:gammaMW}
\end{figure}

In the analysis described so far, we have assumed $\gamma_{\rm MW}=1.0$. If a smaller inner slope were chosen, we would expect an increase in the best-fit annihilation cross-section for the signal. This would lead to correspondingly more stringent constraints on the dwarf spheroidals. We therefore repeat the calculation for several values of $\gamma_{\rm MW}$, as well as for Einasto profiles with parameter $\alpha_{\rm MW}$ to produce constraints in the $\gamma_{\rm dsph}-\gamma_{\rm MW}$ plane and $\alpha_{\rm dsph}-\alpha_{\rm MW}$ plane, shown in figure~\ref{fig:gammaMW}.

\section{Comparison To Simulations and Observation}
\label{sec:comparison}

It has long been suspected that there is a discrepancy between the observed and simulated profiles of dwarf galaxies. For a review of observational evidence and evidence from numerical simulations, see~\cite{deBlok:2009sp}. Early attempts to fit the observational data to an analytic profile~\cite{Moore:1994yx,Flores:1994gz} showed that dwarf galaxies are well characterized as having a constant density core ($\gamma=0$) following an isothermal profile:
\begin{align}
\rho_I=\frac{\rho_0}{1+(r/R_C)^2},
\end{align}
where $\rho_0$ is the central density and $R_C$ is the core radius.
A variation on the isothermal profile, the Burkert profile~\cite{Burkert:1995yz} was later introduced to account for observations indicating that the density falls off as $r^{-3}$ at large radii:
\begin{align}
\rho_B=\frac{\rho_0}{\left(1+r/R_C\right)\left(1+(r/R_C)^2\right)}.
\end{align}
Numerous other groups have found evidence for cored, rather than cuspy, halos in dwarf galaxies~\cite{Burkert:1997td,vandenBosch:2000rza,Salucci:2002nc,Gentile:2006hv}

Few studies present a numerical best fit value for the inner slope, instead typically comparing the NFW ($\gamma=1$) model to an isothermal or Burkert profile ($\gamma=0$). Those that do (several examples of which are listed below) tend to find values of $\gamma\sim0.2$. In their measurement of the dwarf irregular galaxy NGC 6822, ref.~\cite{Weldrake:2002ri} finds an inner slope between $\gamma=0.13\pm0.14$ to $\gamma=-0.22\pm0.35$ depending on the resolution chosen. Spekkens et al.~\cite{Spekkens:2005ik} have derived density profiles for 165 low-mass galaxies including dwarf galaxies based on their rotation curves to find median inner slopes of $\gamma=0.22\pm0.08$ to $0.28\pm0.06$ depending on the subsample considered.

Numerical simulations of cold dark matter (CDM) halos, on the other hand, have typically found values of the inner slope greater than $\gamma=1$. Early numerical simulations of CDM halos were well characterized by the NFW profile of equation (\ref{eq:NFW}) with $\gamma\sim 1$~\cite{Dubinski:1991bm,Navarro:1996gj,Colin:2003jd} for halos of all sizes. Others pointed towards an even steeper slope of $\gamma\sim 1.5$\cite{Moore:1999gc,Klypin:2000hk} or an intermediate value of $\gamma\sim 1.2$~\cite{Diemand:2005wv}. Despite this variation, there is general agreement that pure CDM simulations result in inner slopes of $\gamma\geq 1$.

Some simulations instead found that the slope continues to become more shallow at smaller radii but does not converge\cite{Navarro:2003ew,Hayashi:2003sj}. The Einasto profile, eq. (\ref{eq:Einasto})\cite{Merritt:2005xc,Graham:2006ae}, parameterizes this kind of behaviour. It describes a cored profile at large values of $\alpha$ and becomes cuspier for small values of order $0.1$. Ref.~\cite{Navarro:2008kc} found CDM simulations are well described by $\alpha\approx 0.17$, which even at $r/r_s=10^{-3}$ provide a slope of $\gamma\sim 1$, and therefore for our purposes represents a cuspy profile.

It is clear that our results for the inner slopes of dwarf spheroidal halos, while compatible with observation, are not compatible with traditional CDM simulations. Our results favour values of $\gamma_{\rm dsph}<1.0$. They also favour $\gamma_{\rm dsph}<\gamma_{\rm MW}$, which would suggest that the inner slope of the Milky Way's profile is steeper than that of dwarf spheroidals.

\section{Beyond CDM}
\label{sec:beyondcdm}

The core/cusp controversy is by no means new, and~\cite{Weinberg:2013aya} reviews it in great detail. Many mechanisms have been proposed through which baryonic matter can have a feedback effect on the dark matter halo in the hopes of giving a more cored halo, but the results have been mixed. These mechanisms include rotating bars\cite{Weinberg:2001gm} (however later studies argue that this might actually have the opposite effect\cite{Dubinski:2008rn}) and the heating of cusps by dynamical friction\cite{ElZant:2001re,Tonini:2006gwz,RomanoDiaz:2008wz} (however again, others find that this process is insufficient to explain cored profiles\cite{Jardel:2008bi}). Another possibility is feedback from supernovae\cite{Pontzen:2011ty,Governato:2012fa}; in these simulations repeated feedback from supernovae can turn a cusp into a core. Although viable baryonic mechanisms have been proposed to explain the discrepancy, its ultimate source remains an open question.

Although the standard CDM paradigm is difficult to render consistent with cored profiles, some dark matter models address this issue. Models of warm dark matter (WDM) such as sterile neutrinos rely on the particles having large velocities during structure formation, giving them a free-streaming length with a similar scale to galaxies. This smooths out density fluctuations on scales less than the free streaming length, and is borne out in simulations of WDM halos, giving dwarf sized halos a more cored profile\cite{AvilaReese:2000hg,Colin:2007bk,Dunstan:2011bq,Maccio:2012qf,Schneider:2013wwa,Angulo:2013sza}, though WDM still faces some challenges, including conflict with the small scale power spectrum~\cite{Polisensky:2010rw}, tension with strong-lens system observations which show evidence for a larger subhalo population than would be produced by WDM~\cite{Dalal:2001fq}, and challenges from observations of the Lyman-$\alpha$ forest which sets a lower limit on the dark matter mass of a few keV~\cite{Seljak:2006qw,Viel:2007mv}. Regarding the specific values of $\gamma$, ref.~\cite{Gonzalez-Samaniego:2015sfp} compares CDM and WDM simulations and find $\gamma=1.18$-$1.46$ for CDM and $\gamma=0.25$-$0.66$ for WDM.

Another solution to the cusp-core problem is self-interacting dark matter (SIDM), in which cold dark matter has weak-scale interactions or no interactions at all with baryonic matter but a large self-interaction cross section. When the scattering cross section is of the order $\sigma/m_\chi\sim0.1$-$1\,\rm{cm}^2\rm{g}^{-1}$, dark matter halos naturally form cores\cite{Spergel:1999mh,Rocha:2012jg,Peter:2012jh}.

An interesting possibility is that of dark matter self-interacting through a light mediator. This results in a scattering cross section inversely proportional to velocity, causing greater self-interactions in dwarf galaxies than in galaxies or clusters\cite{Loeb:2010gj}. For some choices of parameters, the cross section can be up to 100 times greater at velocities typically found in dwarf galaxies than for larger galaxies, which allows cored profiles to form for dwarfs but not for larger halos.  These results correspond well to those presented here: the dwarf spheroidal halos are constrained to be more cored than that of the Milky Way. This `dark force' scattering can be further enhanced at dwarf-scale velocities by resonances, and the coupling can even be chosen such that the correct relic density is reproduced~\cite{Tulin:2012wi,Tulin:2013teo,Kaplinghat:2015aga}.

As WDM and SIDM are able to create cored halos, our results are consistent with these models which depart from the traditional CDM model. This implies that the GCE, if it does prove to originate from annihilating dark matter, would provide evidence in favour of these non-CDM cosmologies.

~\\
{\bf Acknowledgments}

  I would like to thank James Cline and Jonathan Cornell for their ideas and assistance in preparing and reviewing this work. I also thank Matthew Walker for helpful correspondence and Marla Geha for providing us with the stellar kinematic data for several dwarf galaxies through private correspondence. This work was done with the support of the McGill Space Institute and the Natural Sciences and Engineering Research Council of Canada.

\appendix
\section{Maximum Likelihood Method}
\label{app:likelihood}

We adopt the method of Geringer-Sameth et. al~\cite{Geringer-Sameth:2014yza} to calculate the halo parameters using the maximum likelihood method. They argue that the velocity data sample a Gaussian distribution, and therefore adopt the likelihood~\cite{Geringer-Sameth:2014yza}:
\begin{align}
L&=\prod_{i=1}^N\frac{\exp\left[-\frac{1}{2}\frac{\left(u_i-\langle u\rangle\right)^2}{\delta^2_{u,i}+\sigma^2(R_i)}\right]}{\left(2\pi\right)^{1/2}\left(\delta^2_{u,i}+\sigma^2(R_i)\right)^{1/2}},
\end{align}
where $u_i$ and $\delta_{u,i}$ are the observed line of sight velocity and uncertainty, $\langle u\rangle$ is the mean velocity of the dwarf, and $\sigma^2(R_i)$ is the  velocity dispersion at the projected position of the observed star.

The velocity dispersion is the model dependent quantity, and has the form~\cite{Mamon:2004xk}
\begin{align}
\sigma^2(R)&=\frac{2G}{\Sigma(R)}\int_R^\infty \frac{v(s)M(s)}{s^2}\sqrt{s^2-R^2}ds\\
\end{align}
for an isotropic halo. $M(r)$ is the mass contained within the given radius, and $v(r)$ and $\Sigma(R)$ are the stellar density and luminosity profiles respectively. 

For a halo in which stars are distributed according to a Plummer profile~\cite{Geringer-Sameth:2014yza}, the ratio of these profiles is given by
\begin{align}
\frac{v(r)}{\Sigma(R)}&=\frac{3}{4r_{1/2}}\frac{1}{\sqrt{1+r^2/r_{1/2}^2}},\\
\end{align}
where $r_{1/2}$ is the half-light radius.

\begin{table*}[p]

\begin{tabular}{|l|cc|cc|cc|cc|cc|c|}
\hline
Dwarf Galaxy&\multicolumn{2}{|c|}{$\gamma=0.2$}&\multicolumn{2}{|c|}{$\gamma=0.4$}&\multicolumn{2}{|c|}{$\gamma=0.6$}&\multicolumn{2}{|c|}{$\gamma=0.8$}&\multicolumn{2}{|c|}{$\gamma=1.0$ $^b$}&Ref.\\
&$R_s$ [kpc]&$\rho_s$ [GeV/cm$^3$]&$R_s$&$\rho_s$&$R_s$&$\rho_s$&$R_s$&$\rho_s$&$R_s$&$\rho_s$&\\
\hline
Carina&$0.68$&$2.4$&$0.79$&$1.6$&$0.93$&$1.0$&$1.1$&$0.6$&$1.4$&$0.32$&\cite{Walker:2008ax}\\
Draco&$1.4$&$4.7$&$1.8$&$2.7$&$2.6$&$1.4$&$4.6$&$0.49$&$-$&$-$&\cite{Walker:2015}\\
Fornax&$0.66$&$6.6$&$0.74$&$4.8$&$0.84$&$3.3$&$0.98$&$2.1$&$1.2$&$1.3$&\cite{Walker:2008ax}\\
Leo I&$1.1$&$3.4$&$1.4$&$2.1$&$1.9$&$1.2$&$2.8$&$0.53$&$5.5$&$0.15$&\cite{Mateo:2007xh}\\
Leo II&$1.1$&$3.1$&$1.4$&$1.9$&$1.9$&$1.1$&$2.8$&$0.48$&$6.1$&$0.13$&\cite{Koch:2007ye}\\
Sculptor&$0.57$&$6.6$&$0.65$&$4.5$&$0.76$&$2.9$&$0.92$&$1.8$&$1.2$&$0.98$&\cite{Walker:2008ax}\\
Sextans&$0.59$&$3.5$&$0.68$&$2.4$&$0.8$&$1.6$&$0.97$&$0.94$&$1.2$&$0.52$&\cite{Walker:2008ax}\\
Bootes I&$1.7$&$6.4$&$2.4$&$3.6$&$3.9$&$1.6$&$36$&$0.13$&$-$&$-$&\cite{Koposov:2011zi}\\
Hercules&$1.0$&$0.62$&$1.2$&$0.39$&$1.6$&$0.22$&$2.2$&$0.11$&$3.7$&$0.039$&\cite{Simon:2007dq}$^a$\\
Leo V&$2.0$&$1.3$&$2.8$&$0.73$&$5.5$&$0.3$&$40$&$0.012$&$-$&$-$&\cite{Walker:2009iq}\\
Segue 1&$1.1$&$4.4$&$1.4$&$2.7$&$1.9$&$1.5$&$2.8$&$0.67$&$6.4$&$0.17$&\cite{Simon:2010ek}\\
Segue 2&$1.4$&$4.9$&$1.8$&$2.9$&$2.6$&$1.5$&$4.9$&$0.54$&$-$&$-$&\cite{Kirby:2013isa}\\
Canes Venatici I&$1.9$&$1.1$&$2.6$&$0.59$&$4.3$&$0.26$&$16$&$0.047$&$-$&$-$&\cite{Simon:2007dq}$^a$\\
Canes Venatici II&$1.5$&$5.1$&$2$&$4.1$&$2.9$&$2$&$6.3$&$0.66$&$-$&$-$&\cite{Simon:2007dq}$^a$\\
Coma Berenices&$1.4$&$6$&$1.9$&$3.5$&$2.7$&$1.8$&$5.4$&$0.62$&$-$&$-$&\cite{Simon:2007dq}$^a$\\
Leo T$^c$&$0.076$&$210$&$0.088$&$140$&$0.1$&$86$&$0.13$&$50$&$0.16$&$27$&\cite{Simon:2007dq}$^a$\\
UrsaMajor I&$0.16$&$30$&$0.18$&$21$&$0.21$&$14$&$0.25$&$8.3$&$0.31$&$4.8$&\cite{Simon:2007dq}$^a$\\
UrsaMajor II&$1.6$&$3.6$&$2.1$&$2.1$&$3.2$&$0.99$&$8$&$0.27$&$-$&$-$&\cite{Simon:2007dq}$^a$\\

\hline
\multicolumn{12}{l}{$^a$ Unpublished, provided by private correspondence.}\\
\multicolumn{12}{l}{$^b$ For missing data, see explanation in text.}\\
\multicolumn{12}{l}{$^c$ Due to lack of FERMI-LAT data, this dwarf is excluded from constraints on $\gamma$.}
\end{tabular}
\caption{Best-fit NFW parameters for various values of $\gamma_{\rm dsph}$.}
\label{tab:NFW}
\end{table*}

\begin{table*}[p]

\begin{tabular}{|l|cc|cc|cc|cc|cc|c|}
\hline
Dwarf Galaxy&\multicolumn{2}{|c|}{$\alpha=0.2$}&\multicolumn{2}{|c|}{$\alpha=0.4$}&\multicolumn{2}{|c|}{$\alpha=0.6$}&\multicolumn{2}{|c|}{$\alpha=0.8$}&\multicolumn{2}{|c|}{$\alpha=1.0$}&Ref.\\
&$R_s$ [kpc]&$\rho_s$ [GeV/cm$^3$]&$R_s$&$\rho_s$&$R_s$&$\rho_s$&$R_s$&$\rho_s$&$R_s$&$\rho_s$&\\
\hline
Carina&$1.6$&$0.061$
&$1.3$&$0.11$
&$1.2$&$0.16$
&$1.2$&$0.20$&
$1.1$&$0.24$&\cite{Walker:2008ax}\\
Draco&$15$&$0.016$&$2.9$&$0.19$&$1.8$&$0.44$&$1.4$&$0.67$&$1.2$&$0.85$&\cite{Walker:2015}\\
Fornax&$1.1$&$0.34$&$1.4$&$0.24$&$1.5$&$0.21$&$1.6$&$0.2$&$1.7$&$0.19$&\cite{Walker:2008ax}\\
Leo I&$8.0$&$0.02$&$2.2$&$0.16$&$1.5$&$0.3$&$1.2$&$0.41$&$1.1$&$0.48$&\cite{Mateo:2007xh}\\
Leo II&$8.2$&$0.018$&$2.2$&$0.14$&$1.5$&$0.27$&$1.2$&$0.36$&$1.1$&$0.42$&\cite{Koch:2007ye}\\
Sculptor&$1.3$&$0.19$&$1.2$&$0.24$&$1.2$&$0.26$&$1.2$&$0.28$&$1.2$&$0.29$&\cite{Walker:2008ax}\\
Sextans&$1.3$&$0.11$&$1.2$&$0.14$&$1.2$&$0.15$&$1.2$&$0.16$&$1.2$&$0.17$&\cite{Walker:2008ax}\\
Bootes I&$38$&$0.013$&$4.1$&$0.24$&$2.1$&$0.59$&$1.6$&$0.88$&$1.3$&$1.1$&\cite{Koposov:2011zi}\\
Hercules&$5.6$&$0.0048$&$1.8$&$0.03$&$1.3$&$0.052$&$1.1$&$0.068$&$1.5$&$0.62$&\cite{Simon:2007dq}$^a$\\
Leo V&$63$&$0.0021$&$5.1$&$0.047$&$2.4$&$0.12$&$1.7$&$0.18$&$1.4$&$0.23$&\cite{Walker:2009iq}\\
Segue 1&$8.7$&$0.025$&$2.2$&$0.2$&$1.4$&$0.38$&$1.2$&$0.51$&$1.1$&$0.58$&\cite{Simon:2010ek}\\
Segue 2&$17$&$0.017$&$3$&$0.2$&$1.7$&$0.52$&$1.4$&$0.69$&$1.2$&$0.8$&\cite{Kirby:2013isa}\\
Canes Venatici I&$40$&$0.002$&$4.5$&$0.039$&$2.3$&$0.1$&$1.7$&$0.16$&$1.4$&$0.2$&\cite{Simon:2007dq}$^a$\\
Canes Venatici II&$22$&$0.021$&$3.3$&$0.28$&$1.8$&$0.48$&$1.4$&$0.65$&$1.3$&$0.76$&\cite{Simon:2007dq}$^a$\\
Coma Berenices&$19$&$0.019$&$3.1$&$0.25$&$1.8$&$0.54$&$1.4$&$0.77$&$1.2$&$0.92$&\cite{Simon:2007dq}$^a$\\
Leo T$^c$&$0.16$&$6.2$&$0.16$&$7.3$&$0.17$&$7.6$&$0.18$&$8.2$&$0.18$&$9.4$&\cite{Simon:2007dq}$^a$\\
UrsaMajor I&$0.32$&$1.1$&$0.35$&$0.96$&$0.41$&$0.72$&$0.47$&$0.56$&$0.52$&$0.46$&\cite{Simon:2007dq}$^a$\\
UrsaMajor II&$26$&$0.0092$&$3.5$&$0.14$&$1.9$&$0.33$&$1.5$&$0.48$&$1.3$&$0.59$&\cite{Simon:2007dq}$^a$\\

\hline
\multicolumn{12}{l}{$^a$ Unpublished, provided by private correspondence.}\\
\multicolumn{12}{l}{$^c$ Due to lack of FERMI-LAT data, this dwarf is excluded from constraints on $\gamma$.}
\end{tabular}
\caption{Best-fit Einasto parameters for various values of $\alpha_{\rm dsph}$.}
\label{tab:Einasto}
\end{table*}

The mass contained within a given radius is attained by integrating the chosen density profile:
\begin{align}
M(s)=\int_0^s4\pi r^2\rho_s(r,R_s,\rho_s,\gamma)dr.
\end{align}

For each dwarf spheroidal we minimize the negative log likelihood for $0.1\leq\gamma\leq1.2$ and again for $0.1\leq\alpha\leq1.0$ (for the NFW and Einasto profiles respectively) over the parameters $R_s$ and $\rho_s$. This is accomplished using the downhill simplex method over the two parameters. The best fit values of $R_s$ and $\rho_s$ are shown in table~\ref{tab:NFW} for several values of $\gamma$ and in table~\ref{tab:Einasto} for $\alpha$, the Einasto profile parameter.

Best fit values are not available for some dwarf galaxies for $\gamma=1.0$ (or greater). The likelihood in these cases approaches its maximum value only as $r_s\rightarrow\infty$ and $\rho_s\rightarrow\infty$. This is due to the nature of the NFW profile, which has its shallowest log slope at $r=0$, with the slope becoming steeper at greater distances. In these cases, therefore, the slope $\gamma=1.0$ is inconsistent with the stellar kinematic data. In these cases the fit can always be made better by increasing $r_s$ to grant a smaller log slope (approaching a uniform log slope of $1.0$) and reducing the density to compensate.

\bibliography{jfactorbib}{}
\bibliographystyle{hep}

\end{document}